# Cryptocurrencies without Proof of Work

Iddo Bentov*
Computer Science Dept., Technion
idddo@cs.technion.ac.il

Ariel Gabizon†
Computer Science Dept., Technion
ariel.gabizon@gmail.com

Alex Mizrahi
chromawallet.com
alex.mizrahi@gmail.com

## Abstract

We study decentralized cryptocurrency protocols in which the participants do not deplete physical scarce resources. Such protocols commonly rely on *Proof of Stake*, i.e., on mechanisms that extend voting power to the stakeholders of the system. We offer analysis of existing protocols that have a substantial amount of popularity. We then present our novel pure *Proof of Stake* protocols, and argue that they help in mitigating problems that the existing protocols exhibit.

## 1 Introduction

The decentralized nature of Bitcoin [12, 20] means that anyone can become a "miner" at any point in time, and thus participate in the security maintenance of the Bitcoin system and be compensated for this work. The miners continuously perform *Proof of Work* (PoW) computations, meaning that they attempt to solve difficult computational tasks. The purpose of the PoW element in the Bitcoin system is to reach consensus regarding the ledger history, thereby synchronizing the transactions and making the users secure against double-spending attacks.

The miners who carry out PoW computations can be viewed as entities who vote on blocks of transactions that the users recently broadcasted to the network, so that the decision-making power of each miner is in proportion to the amount of computational power that she has. Thus, an individual miner who has a fraction $p$ of the total mining power can create each new block with probability $\approx p$, though other factors such as "selfish mining" [2, 10, 11, 29] can influence $p$.

Under the assumption that the majority of the PoW mining power follows the Bitcoin protocol, the users can become increasingly confident that the payment transactions that they receive will not be reversed [12, 20, 26].

By means of the PoW mechanism, each miner depletes physical scarce resources in the form of electricity and mining equipment erosion, and thereby earns cryptographic scarce resources in the form of coins that can be spent within the Bitcoin system.

Hence the following question is of interest: can a *decentralized* cryptocurrency system be as secure as Bitcoin even if the entities who maintain its security do not deplete physical scarce resources?

Cryptocurrency protocols that attempt to avoid wasting physical scarce resources commonly rely on *Proof of Stake*, i.e., on mechanisms that give the decision-making power regarding the continuation of the ledger history to entities who possess coins within the system. The rationale behind *Proof of Stake* is that entities who hold stake in the system are well-suited to maintain its security, since their stake will diminish in value when the security of the system erodes. Therefore, in an analogous manner to Bitcoin, an individual stakeholder who possesses $p$ fraction of the total amount of coins in circulation becomes eligible to create the next extension of the ledger with probability $\approx p$.

*Supported by funding from the European Community's Seventh Framework Programme (FP7/2007-2013) under grant agreement number 240258.

†Supported by funding from the European Community's Seventh Framework Programme (FP7/2007-2013) under grant agreement numbers 257575 and 240258.

We use the terminology "pure" *Proof of Stake* to refer to a cryptocurrency system that relies on *Proof of Stake* and does not make any use of PoW. To the best of our knowledge, the idea of *Proof of Stake* in the context of cryptocurrencies was first introduced in [32], though that discussion focused on non-pure *Proof of Stake* variants.

PoW based cryptocurrencies become insecure when a significant enough portion of the total mining power colludes in an attack. Likewise, the security of pure *Proof of Stake* cryptocurrencies deteriorates when enough stakeholders wish to collude in an attack. If the majority of the stake wishes to participate in attacks on a pure *Proof of Stake* system, it can be argued that there is no longer enough interest that this system should continue to exist, hence assuming that the majority of the stake will not participate in an (overt) attack is sensible. The same does not necessarily hold in a PoW based system, i.e., the majority of the mining power might be under the control of an external adversary during some time period, while the majority of the participants in this system still wish for it to remain sound. See [4] and Section 3 for additional considerations.

## 1.1 Related work

In [4], authors of this work presented a hybrid protocol that relies both on PoW and *Proof of Stake*, where the objective is to combine to advantageous properties of the PoW element and the *Proof of Stake* element into a system that is superior to relying on only one of these two elements.

An alternative to PoW that is based on data storage is presented in [1, 9], though in such *Proof of Space* systems the participants may still opt to make use of PoW due to an inherent time-space tradeoff. Recent progress on *Proof of Space* based cryptocurrency protocols is worth consideration [23, 24]. There are also cryptocurrencies that rely solely on PoW for security, but require the miners to use disk storage for other useful purposes [22]. In any case, *Proof of Space* based protocols still require depletion of physical scarce resources to some significant degree, unlike the focus of this work.

## 1.2 Organization of the paper

The main contributions of this work are organized as follows. In Section 2.1 we analyse an existing pure *Proof of Stake* system. In Section 2.2 we present a novel pure *Proof of Stake* protocol, analyse its properties, and argue that it is more secure than the existing system. In Section 2.3 we present another variant of our pure *Proof of Stake* protocol, and explore the tradeoffs that it offers in terms of security, fairness, and efficiency. In Section 3 we explain the greater importance of checkpointing the ledger history of pure *Proof of Stake* systems compared to PoW based systems, and specify our proposal for handling this issue. In Section 4 we examine how to conduct the initial issuance of coins in a pure *Proof of Stake* system.

# 2 Pure *Proof of Stake*

There are two apparent hurdles with decentralized pure *Proof of Stake* systems: fair initial distribution of the money supply to the interested parties, and network fragility if the nodes are rational rather than altruistic. PoW offers an elegant solution to the first hurdle, by converting physical scarce resources into coins in the system. We provide here an analysis of the second hurdle in an existing pure *Proof of Stake* system, and also describe our novel CoA and Dense-CoA pure *Proof of Stake* systems that seek to mitigate this problem. Let us note that the second hurdle is less severe in PoW systems, though bribe attacks on Bitcoin have indeed been considered, for example in [30].

## 2.1 The PPCoin system

PPCoin is a pure *Proof of Stake* system, in the sense that PoW is used only for distributing the initial money supply[1]. Stakeholders in the PPCoin network can create the next block according to the following type of condition:

$$\text{hash}(\texttt{prev\_blocks\_data}, \texttt{time\_in\_seconds}, txout_A) \leq d_0 \cdot \texttt{coins}(txout_A) \cdot \texttt{timeweight}(txout_A) \quad (*)$$

In the inequality (*), `time_in_seconds` should correspond to the current time (with some leniency bounds), thus restricting hash attempts to 1 per second and preventing PoW use at creating the next block, because nodes will regard a new block as invalid unless the difference between its time and their local time is within the bounds. The notation $\texttt{coins}(txout_A)$ refers to the amount of coins of some unspent transaction output $txout_A$, hence if stakeholder $A$ has the private key $sk_A$ that controls $txout_A$ then she can create a valid block by signing the block with $sk_A$ and attaching the signature as evidence that condition (*) holds. This means that a stakeholder who controls an output of e.g. 50 coins is 10 times more likely to create a block than a stakeholder who controls an output of 5 coins. See Section 2.1.3 regarding $\texttt{timeweight}(txout_A)$, and Section 2.1.4 regarding `prev_blocks_data`. The constant $d_0$ is readjusted according to a protocol rule that dictates that blocks should be created in intervals of 10 minutes on average, i.e., if fewer stakeholders are online during a certain time period then $d_0$ gets increased. The winning blockchain is the one with the largest cumulative stake, i.e., the blockchain with the most blocks such that stake blocks are weighted according to their $d_0$ difficulties, and PoW blocks have a negligible weight.

Although the PPCoin cryptocurrency had a market cap of over $100 million in 2014, the PPCoin protocol has the following problems:

### 2.1.1 Rational forks

On every second we have that $\Pr[\{\text{some block is solved}\}] \approx \frac{1}{600}$, therefore multiple blocks will be solved simultaneously every $\approx 360000$ seconds $\approx 4$ days. Rational stakeholders can increase their expected reward by maintaining and trying to solve blocks on the multiple forked chains that were transmitted to them, which would lead to a divergent network. An individual stakeholder can either tie her hands behind her back by ignoring all the forked chains except for one, or opt to gain more rewards by keeping all the forked chains, which may render her entire stake worthless in case the network becomes divergent. The strategy of tying your hands behind your back is not a Nash equilibrium: if all the stakeholders follow this strategy then it is better for an individual stakeholder to deviate and maintain all the forked chains, as her influence on the overall convergence of the network is minor. Network propagation lag implies an even greater frequency of forks, as a stakeholder will get competing blocks sent to her even if those blocks were honestly solved a few seconds apart from one another. Worse still, when a rational stakeholder who currently tries to extend the block $B_i$ receives $B_{i+1}$ from her peers, she may opt to increase her expected reward by attempting to extend both the chain $\ldots, B_i, B_{i+1}$ and the chain $\ldots, B_i$ simultaneously. Rational stakeholders may thus prefer to reject blocks whose timestamp is later than another block that they currently try to extend, though an attempt to extend both $\ldots, B_i, B_{i+1}$ and $\ldots, B_i$ can still be possible if the rule that the stakeholders deploy does not retrace to an earlier chain that is received late due to propagation lag.

### 2.1.2 Bribe attacks on PPCoin

An attacker can double-spend quite easily. After the merchant waits for e.g. 6 block confirmations and sends the goods, the attacker can publicly announce her intent to create a fork that reverses the last 6 blocks, and offer bribes to stakeholders who would sign blocks of her competing branch that starts 6 blocks earlier. The attacker may offer a larger bribe to stakeholders who sign only her branch, and may commit to giving bribes even after her competing branch wins, to encourage more stakeholders to participate in the attack. Notice that the stakeholders who collude with the attacker will not lose anything in case the attack fails. As long as the value of the goods is greater than the total value of the bribes, this attack will be profitable. Let us note that a bribe attack in a pure PoW network has to surmount far greater obstacles: miners who join the attack would deplete their resources while working on a fork with a 6 blocks deficit, and it is a nontrivial task to assess the success probability by measuring how many other miners participate in the attack. See also [4, Section 5.3].

---

[1] See <http://peercoin.net/assets/paper/peercoin-paper.pdf>. It should be noted that the PPCoin protocol keeps being tweaked, which can easily be done because it is controlled by a centralized entity via signed checkpoints, as opposed to being controlled by a decentralized network. For security, the users of the PPCoin system currently rely on PoW blocks and on the checkpoints of the centralized entity.

### 2.1.3 Attacks associated with accumulation of timeweight

In PPCoin protocol $v0.2$, `timeweight`($txout_A$) is proportional to the time elapsed since the transaction whose output is $txout_A$ was included into a block, and is not capped. The meaning of this is that the probability to generate a block immediately after a stake was used is very low, but it grows as time passes. This avoids the exponential distribution between payouts by boosting the chances of stakeholders who possess a small amount of coins to generate a block in a reasonable time frame, while large stakeholders will still be eligible to create blocks without much wait. Since there is no good way to have stake pools, the `timeweight` component is particularly useful.

However, this also means that an attacker can boost her chances to generate consecutive blocks by waiting. For example, if an attacker controls 10% of all the coins, and she waits until the average `timeweight` of the transaction outputs she controls is 5 times higher than the average `timeweight` of transaction outputs that belong to the other stakeholders who participate, she will be able to produce a chain of multiple consecutive blocks, as the probability that the next block is produced by her is close to 50% under these conditions.

The way by which `timeweight` is computed was changed in PPCoin $v0.3$, and in this protocol version `timeweight` stops growing after 90 days. This means that under the condition that the average `timeweight` is close to the cap $60 \cdot 60 \cdot 60$, an attacker will get no advantage by waiting. This condition is met when total number of participating transaction outputs is higher than $2 \cdot 90 \cdot 60 \cdot 60$, and in that case most unspent transaction outputs will wait for more than 90 days until they generate a stake.

Hence, this attack is now considerably less effective, but the beneficial properties that `timeweight` was designed to achieve are diminished.

### 2.1.4 Opportunistic attacks in relation to the need to disallow PoW

A stakeholder who holds a significant fraction of all the coins is able to generate a significant fraction of the blocks, as the probability to generate a block is proportional to the amount of coins that a stakeholder holds. Therefore, from time to time a stakeholder will be able to generate chains of consecutive blocks.

We can analyze this event by using a simplified model where stakeholders who own $\frac{1}{M}$ of all coins can generate a block with probability $\frac{1}{M}$, and the probability to generate $k$ sequential blocks is $(\frac{1}{M})^k$. This approximation is accurate under the assumption that the stakeholder holds a number of unspent transaction outputs significantly larger than $k$, so that `timeweight` will have no impact. We can estimate the average number of blocks between groups of $k$ sequential blocks generated by one stakeholder as a mean of exponential distribution, which would be equal to $1/(1/M)^k = M^k$.

If merchants wait for $k$ confirmations before sending their goods, the stakeholder has a chance to attack the merchant when she is able to generate $k$ sequential blocks, thus the mean number of blocks between such attacks is $M^k$. For example, a stakeholder who holds $\frac{1}{4}$ of all coins participating in stake mining will be able to carry out a 6-block reorganization each $4^6 = 4096$ blocks, i.e., approximately once per month if one block is generated every 10 minutes.

An attacker who is able to create $k$ sequential blocks would prefer to know about it as early as possible, so that she has enough time to send the payment transaction (that she intends to reverse) to the merchant. If the possible stakeholders' identities who may create the next blocks are derived from a low entropy process that only takes into account the identities who created the previous blocks, then the attacker can "look into the future" by carrying out brute-force computations to assess the probabilities that she will be able to create the $k$ consecutive blocks at certain points in time. In order to gain a measure of unpredictability, PPCoin re-calculates once every 6 hours a "stake modifier" value that depends on the transactions that the previous blocks included, i.e., this stake modifier is part of `prev_blocks_data` in condition (*). Therefore, a stakeholder who obtains an opportunity to generate $k$ blocks in a row can know about this approximately 6 hours in advance, so she has plenty of time to mount an attack. If the protocol required the stake modifier to be re-calculated at a shorter time interval, this would open the door wider for a rational stakeholder to do PoW attempts at deriving herself as eligible to create future blocks more frequently.

## 2.2 The CoA pure *Proof of Stake* system

The Chains of Activity (CoA) system that we hereby present is a pure *Proof of Stake* protocol that aims to overcome the problem of rational forks (cf. Section 2.1.1) by dictating that only a single stakeholder identity may create the next block, and solidifying the random choices for these identities in the earlier ledger history via an interleaving mechanism.

The CoA protocol is based in part on the core element of PoA [4], i.e., on a lottery among the online stakeholders via the *follow-the-satoshi* procedure. This procedure takes as input an index of a satoshi (smallest unit of the cryptocurrency) between zero and the total number of satoshis in circulation, fetches the block of ledger data in which this satoshi was minted, and tracks the transactions that moved this satoshi to subsequent addresses until finding the stakeholder who can currently spend this satoshi (cf. [4, Section 3 and Appendix A]). Note that if for example Alice has 6 coins and Bob has 2 coins then Alice is 3 times more likely to be picked by *follow-the-satoshi*, regardless of how their coins are fragmented. This implies that a stakeholder who holds her coins in many Sybil addresses do not obtain any advantage with regard to *follow-the-satoshi*.

The CoA protocol is parameterized by an amount of minted satoshis $2^\kappa$, a subgroup length $w \geq 1$, a group length $\ell = \kappa \cdot w$, a function $\mathbf{comb} : \{0,1\}^\ell \to \{0,1\}^\kappa$, a minimal block interval time $G_0$, a minimal stake amount $C_0$, an award amount $C_1$ such that $0 \leq C_1 < C_0$, and a double-spending safety bound $T_0$.

The blocks creation process of CoA assembles a blockchain that is comprised of groups of $\ell$ consecutive blocks:

$$\overbrace{\square\square\cdots\square}^{\ell}, \overbrace{\square\square\cdots\square}^{\ell}, \overbrace{\square\square\cdots\square}^{\ell}, \ldots$$

The rules of the CoA protocol are specified as follows:

### The CoA Protocol

1. Each block is generated by a single stakeholder, whose identity is fixed and publicly known (as will be explained in the next steps). This stakeholder collects transactions that are broadcasted over the CoA network as she sees fit, and then creates a block $B_i$ that consists of these transactions, the hash of the previous block, the current timestamp, the index $i$, and a signature of these pieces of data as computed with her private key.

2. Every newly created block $B_i$ is associated with a supposedly uniformly distributed bit $b_i$ that is derived in a deterministic fashion, for example by taking the first bit of $hash(B_i)$.

3. The time gap between $B_i$ and $B_j$ must be at least $|j - i - 1| \cdot G_0$. This means that if for example the next four blocks $B_i, B_{i+1}, B_{i+2}, B_{i+3}$ were supposed to be generated by the four stakeholders $A_i, A_{i+1}, A_{i+2}, A_{i+3}$ but $A_{i+1}$ and $A_{i+2}$ were inactive, then the difference between the timestamp of $B_{i+3}$ and $B_i$ must be at least $2G_0$. Nodes in the network will consider a newly created block to be invalid if its timestamp is too far into the future relative to their local time.

4. After a group of $\ell$ valid blocks $B_{i_1}, B_{i_2}, \ldots, B_{i_\ell}$ is created, the network nodes will form a $\kappa$-bit seed $S^{B_{i_\ell}} = \mathbf{comb}(b_{i_1}, \ldots, b_{i_\ell})$. The function $\mathbf{comb}$ can simply concatenate its inputs (if $w = 1$), and several other alternatives are explored in Section 2.2.1.

5. The seed $S^{B_{i_\ell}}$ is then used in an interleaved fashion to derive the identities of the *after* next $\ell$ stakeholders, via *follow-the-satoshi*. That is, if the next $\ell$ valid blocks are $B_{i_\ell+j_1}, B_{i_\ell+j_2}, \ldots, B_{i_\ell+j_\ell}$, then the nodes who follow the protocol will derive the identity of the stakeholder who should create the block $B_{i_\ell+j_\ell+z}$ by invoking *follow-the-satoshi* with $hash(i_\ell, z, S^{B_{i_\ell}})$ as input, for $z \in \{1, 2, \ldots\}$.

6. If the derived satoshi is part of an unspent output of $c < C_0$ coins, the stakeholder must also attach an auxiliary signature that proves that she controls another output of at least $C_0 - c$ coins, or else she will not be able to create a valid block. Neither the derived output nor this auxiliary output may be spent in the first $T_0$ blocks that extend the newly created block. In case the stakeholder $A_i$ who should create the $i^{\text{th}}$ block signs two different blocks $B_i, B'_i$, any stakeholder $A_j$ among the next $T_0$ derived stakeholders can include it as evidence in the block that she creates, in order to confiscate at least $C_0$ coins that $A_i$ possessed. The stakeholder $A_j$ is awarded with $C_1$ of the confiscated coins, and the rest of the confiscated coins are destroyed.

7. "Three strikes" blacklisting rule: in case an output $txout_0$ was eligible to create a block three consecutive times but the stakeholder who controls $txout_0$ did not create a block, $txout_0$ becomes blacklisted for the purpose of creating future blocks. That is, starting from the after next group of $\ell$ stakeholders, if *follow-the-satoshi* with $hash(i_\ell, z, S^{B_{i_\ell}})$ leads to $txout_0$, then all the honest nodes who follow the protocol will skip from $z - 1$ directly to $z + 1$. In case $txout_0$ is spent via a regular transaction, the satoshis that $txout_0$ was comprised of are no longer blacklisted.

8. If the network nodes see multiple competing blockchains, they consider the blockchain that consists of the largest number of blocks to be the winning blockchain.

The interleaving in step 5 is crucial as a cementing mechanism. Otherwise, competing last stakeholders may extend the chain with seeds that derive different $\ell$ next identities, introducing divergence risk because it is rational for the next identities to extend the different forks. This cementing process ensures that unless $\approx \ell$ stakeholders collude by bypassing their turn on the honest chain and creating a hidden fork instead, only a single stakeholder will be eligible to create each next block. Thus the rational forks hazard is avoided.

The punishment scheme in step 6 expires after $T_0$ blocks, because honest stakeholder must eventually regain control over their security deposit (refer also to Section 3). Note that a stakeholder can divide her coins among multiple outputs, so that only one of the outputs would become unspendable for $T_0$ blocks. If $C_1 \approx C_0$, an attacker might double-sign and publish the double-signing evidence in a next block to recover her security deposit, so $C_1 \leq \frac{C_0}{2}$ is a better choice.

Let us emphasize that CoA does not need to employ a rule that freezes the output with the winning satoshi of each eligible stakeholder during the entire subchain of $\ell$ blocks. This kind of a rule neither desirable nor necessary. It is undesirable because an unsolicited freezing of funds would make the entire system less attractive, due to the time-value of money. Instead, a CoA stakeholder has to freeze some of her funds only if she wishes to participate in the blocks creation process and earn a reward. The only reason for an unsolicited freezing rule is that it prevents a colluding stakeholder from transferring her coins to the attacker at an early time, and instead requires her to be online and provide her signature as the hostile alternative chain is being built – the distinction between these two attack modes is rather minor.

If $\ell$ is very large (in the extreme $\ell = \infty$, i.e., practically equivalent to selecting the identities of the stakeholders via a round-robin), then an attacker may try to gain possession of future consecutive satoshis in order to mount a double-spending attack (cf. Section 2.2.3). On the other hand, small $\ell$ makes it easier for coalitions to influence the future identities (cf. Section 2.2.1). Moreover, if the range of **comb** were $\kappa' < \kappa$, an attacker could more easily see into the future, e.g. with $\kappa' = 10$ the attacker could buy satoshis of consecutive identities in one possible next group and succeed with probability $1/1024$ to carry out a double-spending attack. A sensible recommendation for the CoA parameters can be $\kappa = 51$ (for $\approx 21$ million coins of $10^8$ satoshis each), $w = 9$ with **comb** as the iterated majority function (see Section 2.2.1), $\ell = 459$, $G_0 = 5$ minutes, and $T_0 = 5000$. It may also make sense to readjust some of the CoA protocol parameters dynamically (cf. [27]).

From the point of view of the users, the performance of the CoA network is better when the interval between blocks is shorter, as this implies that their transactions become confirmed more quickly. We can indeed expect that blocks will be created in intervals of less than $G_0$ minutes in the common case: each eligible stakeholder would wish to earn fees by collecting transactions nearly until the next $G_0$ tick, thereby evading the risk that the next stakeholder will extend an earlier block. Compared to the average 10 minutes interval between blocks of Bitcoin, it is quite reasonable to set $G_0$ to a lower value such as $G_0 = 5$, because the block creators do not race against each other and therefore network connectivity differences have a less pronounced effect (cf. [18, Section 4]).

The purpose of the "three strikes" rule (7) is to avoid the risk of a decreased performance in case a significant amount of stake becomes destroyed. For example, if half of the total stake got destroyed then the performance level would correspond to $2G_0$. Thus, in the unfortunate circumstance that a stakeholder lost her private keys so that neither she nor anyone else has control over the coins that she held, each of the outputs that she controlled will become blacklisted after three strikes. This allows the CoA network to accommodate lost stake while still maintaining the same level of $G_0$ performance. In case a stakeholder finds the private keys that she believed that she lost, she can make a regular transaction and thereby revive her eligibility to create future blocks. The CoA protocol can set any number $s$ as the number of "strikes", but $s = 1$ may result in rather frequent blacklist updates. Hence, $s = 3$ is quite reasonable w.r.t. blacklist maintenance complexity.

Notice that in comparison to Bitcoin, CoA is less susceptible to selfish-mining [11, 29] attacks. This is because one is not destined to earn more rewards by denying others from creating blocks, since the identities of block creators in the near future are already fixed. The CoA protocol can hold an advantage in this regard (over PoW-based protocols) even if the reward is derived only from transaction fees, per Section 2.2.2. This can also be inferred by observing that CoA has no difficulty readjustment rule that an attacker can exploit over time (an honest stakeholder can always revive her eligibility as noted above), while such a rule is the reason why the block creation process in Bitcoin is not memoryless and therefore potentially vulnerable to selfish mining.

The aforementioned concern regarding $\ell = \infty$ can be restated as follows: to have a cryptocurrency protocol that is resistant to double-spending attacks, with no need for PoW and with potentially malicious stakeholders who distrust each other, the only assumptions needed are (1) that the stakeholder who possesses the $i^{\text{th}}$ coin will be online to sign a block when her round (of e.g. 5 minutes) is due, and (2) that the punishment for a dishonest stakeholder who exposes herself by signing two versions of her block in two competing chains is severe enough to prevent the stakeholder from doing that. Alas, assumption (1) is unrealistic, and assumption (2) is nontrivial as it depends on how much stake and anonymity the dishonest stakeholder loses relative to the value of the payment that is being reversed via double-spending. Additionally, any decentralized cryptocurrency system could be forked by some majority group to follow different protocol rules, hence it is crucial to design a protocol such that in equilibrium the protocol benefits the participants of the cryptocurrency.

### 2.2.1 Using low-influence functions to improve chain selection

We compare different ways in which the stakeholders $A_1, \ldots, A_\ell$ who create the current chain segment can choose the stakeholders who will create the next (interleaved) segment. The basic framework is the following.

- At round $i$, the stakeholder $A_i$ will generate and then publish a bit $b_i$ that is supposed to be uniformly distributed.

- The $\ell$ stakeholders of the next chain – denoted $A'_1, \ldots, A'_\ell$ – will be chosen as follows. $A'_i$ will be determined by applying *follow-the-satoshi* on $hash(i_0 + i, \mathbf{comb}(b_1, \ldots, b_\ell))$, where $i_0$ is the index of the stakeholder who precedes $A'_1$. Here $\mathbf{comb} : \{0,1\}^\ell \to \{0,1\}^\kappa$ will be some publicly known function, such that $2^\kappa$ is at least as large as the number of coins in the system.

Denote $s \triangleq \mathbf{comb}(b_1, \ldots, b_\ell)$, i.e., $s$ is the "seed" with which we choose the next set of stakeholders. The main issue is how to choose the function $\mathbf{comb}$ so that a colluding subset of $\{A_1, \ldots, A_\ell\}$ will have a low probability of unfairly influencing the choice of $s$, and therefore have low probability of influencing the choice of $\{A'_1, \ldots, A'_\ell\}$. This is quite similar to the problem of collective sampling considered by Russell and Zuckerman [28] which arises in the context of well-studied questions in cryptography and distributed computing such as leader election and collective coin-flipping (cf. [3]). In fact, we could have used the function **comb** from their 'collective sampling protocol', but do not do so for two reasons.

1. Their protocol requires a rather involved construction of a 'hitting set for combinatorial rectangles' that might be hard to implement in practice in an efficient way.

2. Their construction is not tailored to work well for the ranges of parameters we will consider – when the number of values of the seed – $2^\kappa$ – is 'almost exponential but not exponential' in the number of participants – i.e., $2^\kappa = 2^{\gamma \cdot \ell}$ for constant $\gamma < 1$.

We proceed to give a few choices for the function **comb**. To give an intuitive illustration of the advantages of different choices, we focus on the prominent case of analyzing the probability that the last stakeholder in the chain, $A_\ell$, can choose herself again as one of the first possible stakeholders $A'_1, \ldots, A'_\ell$ of the next round (refer to Figure 1). Denote this probability by $\mu$. We also make the simplifying assumptions that the previous players have indeed picked their bits $b_i$ randomly, and that the function $hash$ is a random oracle. Let us assume that $A_\ell$ has a $q$-fraction of the coins in the system, and denote $p = 1 - (1-q)^\ell$. Thus, $\mu = p$ in case $A_\ell$ picks a random bit.

**Simple concatenation:** We let $\mathbf{comb}(b_1, b_2, \ldots, b_\ell) \triangleq b_1 \circ b_2 \circ \cdots \circ b_\ell$, where $b_i$ is the supposedly random bit that stakeholder $A_i$ provided. The probability $\mu$ that $A_\ell$ can choose herself in the next round is the probability that $\exists b' \in \{0,1\}, i \in \{1, \ldots, \ell\}$ such that $hash(i_0 + i, \mathbf{comb}(b_1, \ldots, b_{\ell-1}, b'))$ maps to a coin of $A_\ell$ under *follow-the-satoshi*. Using the simplifying assumption that these are random independent values we have $\mu = 1 - (1-p)^2 = 2p - p^2 \approx 2p$.

**Combining majority with concatenation:** Assume now that $\ell = \kappa \cdot w$ for positive integer $w$. We now split the $\ell$ stakeholders into groups of size $w$:
$A_1, \ldots, A_w, A_{w+1}, \ldots, A_{2w}, \ldots, A_{(\kappa-1)\cdot w+1}, \ldots, A_{\kappa \cdot w} = A_\ell$. Each group will determine a bit of the seed using the majority function. That is, the $i^{\text{th}}$ bit of the seed, denoted $s_i$, will be the majority of the bits $b_{(i-1)\cdot w+1}, \ldots, b_{iw}$. And $s = \mathbf{comb}(b_1, \ldots, b_\ell) \triangleq s_1 \circ s_2 \cdots \circ s_\kappa$. Note first that when the bits $b_i$ are all chosen randomly, $s$ is random – as the majority of random inputs is a random bit. Now, we analyze again the probability $\mu$ that $A_\ell$ can choose herself in the next round. It can be shown, using Stirling's approximation, that with probability roughly[2] $1 - \sqrt{2/\pi w}$ the last bit of the seed, $s_\kappa$, will already be determined by the bits of the previous stakeholders. This is because when $w$ players choose a bit randomly, the probability that *exactly half* of the bits came out one tends to $\binom{w}{w/2}/2^w \approx \frac{2^{w+1/2}}{\sqrt{\pi w}}/2^w = \sqrt{2/\pi w}$. In the absence of this event we have $\mu = p$, "as it should be". When this event happens, as before $A_\ell$ can get to probability $\approx 2p$. In total we have $\mu \approx p \cdot (1 - \sqrt{2/\pi w}) + 2p \cdot \sqrt{2/\pi w}$. Taking a large enough $w$, this is much closer to the "correct" probability $p$ than in the previous choice of **comb**.

**Improving further by using low-influence functions:** There were two reasons why the majority function was useful above. First, on a random set of inputs, the majority function returns a random bit. In other words, it is a *balanced* function. Second, when all but one of the $w$ inputs are chosen randomly, the probability that the function is not yet determined is low – $\theta(1/\sqrt{w})$. Are there balanced functions where this probability is even lower? If so, we could use them to get an even better construction of the function **comb**. It turns out that this is a very well-known question about the "influence" of boolean functions initially raised in [3], and studied in the seminal paper of Kahn, Kalai and Linial [15] where it was shown that the probability above will always be at least $\Omega(\log w/w)$. Ben-Or and Linial, in fact gave an example of a function where this probability is that low: The TRIBES-function where we divide the $w$ players into "tribes" of size approximately $\log w$, take the AND of the bits in each group, and then take the OR of all resulting bits. However, TRIBES is not known to have efficient deterministic constructions.

---

[2] More precisely, as $w$ goes to infinity this is the limit of the probability of the event.

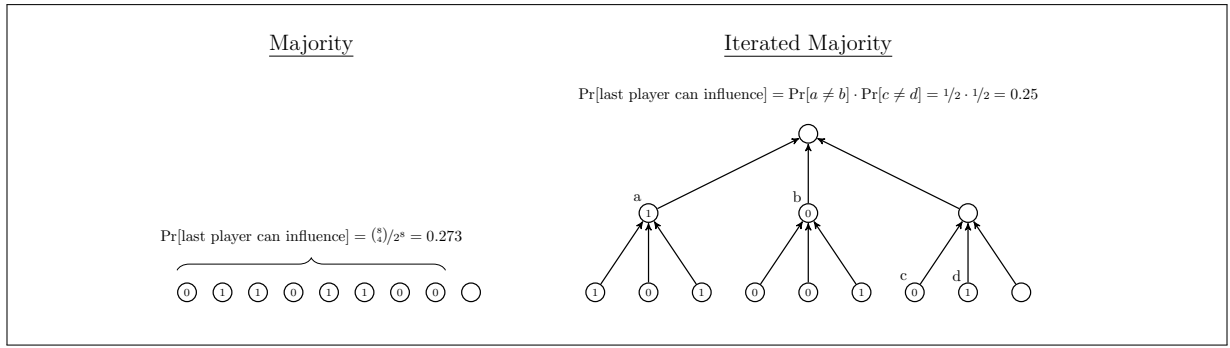

Figure 1: Majority versus Iterated Majority.

**Protection against larger coalitions:** We now address the scenario of a coalition $C \subseteq \{A_1, \ldots, A_\ell\}$ of $|C| = c > 1$ players (i.e., stakeholders) trying to influence the choice of the output of **comb**. For simplicity, we make the following assumptions.

- As before, the $\ell - c$ honest players outside of the coalition choose their input bit randomly.

- We make the worst-case assumption that the $c$ players in the coalition see all the input bits of the honest players before choosing theirs.

- We measure how "successful" a choice for the function **comb** is by the *statistical distance* of the resulting seed to the uniform distribution: For constant $\varepsilon > 0$, a distribution $P$ on $\{0,1\}^\kappa$ is $\varepsilon$-*close to uniform* if for any set $T \subseteq \{0,1\}^\kappa$, the probability that $P \in T$ is at most $|T|/2^\kappa + \varepsilon$. In words, the probability of any event grows at most by an additive factor of $\varepsilon$ compared to it's 'correct' probability.

It turns out that under these assumptions, the problem we are trying to solve is exactly that of constructing *extractors for non-oblivious bit-fixing sources*, considered by Kamp and Zuckerman [16].

Let us use the terminology that a function **comb** : $\{0,1\}^\ell \to \{0,1\}^\kappa$ is an $\varepsilon$-*extractor* if for any choice of the coalition $C$ of size $c$, and any strategy of $C$ to choose their input bits after seeing the bits of the honest players, **comb**$(b_1, \ldots, b_\ell)$ produces an output that is $\varepsilon$-close to uniform.

[16] give the following construction of an $\varepsilon$-extractor – that is in fact the same one we described earlier when replacing the majority function with the *iterated majority* function (defined in [3] and illustrated in Figure 1).

$KZ(b_1, \ldots, b_\ell)$:

- Choose $w = 3 \cdot (c/\varepsilon)^{1/\alpha}$, where $\alpha = \log_3 2$. Set $\ell = w \cdot \kappa$.

- Output $\kappa$ bits by taking the *iterated majority* of consecutive groups of $w$ inputs.

[16], using the analysis of [3] of the iterated majority function, were able to show that the function $KZ$ is an $\varepsilon$-extractor. Upon fixing $\varepsilon$ as the desired statistical error, $\kappa$ as the desired output length, and $\ell$ as the total number of players in a chain, $KZ$ can handle a coalition of size $c \leq \varepsilon \cdot (1/3 \cdot \ell/\kappa)^\alpha$.

On the other hand, [16] also show that any such $\varepsilon$-extractor can handle coalitions of size at most $c \leq \varepsilon \cdot 10 \cdot \ell/(\kappa - 1)$.

Since $\alpha > 1/2$, it follows that this choice of **comb** is less than quadratically worse than the optimal choice. Notice that this assumes that stakeholders who are not honest are non-oblivious, i.e., that they see the choices of the honest stakeholders before they play. This conservative assumption makes a certain sense in our context, as it easier for stakeholders who play in the last locations to try to collude in order to influence the seed.

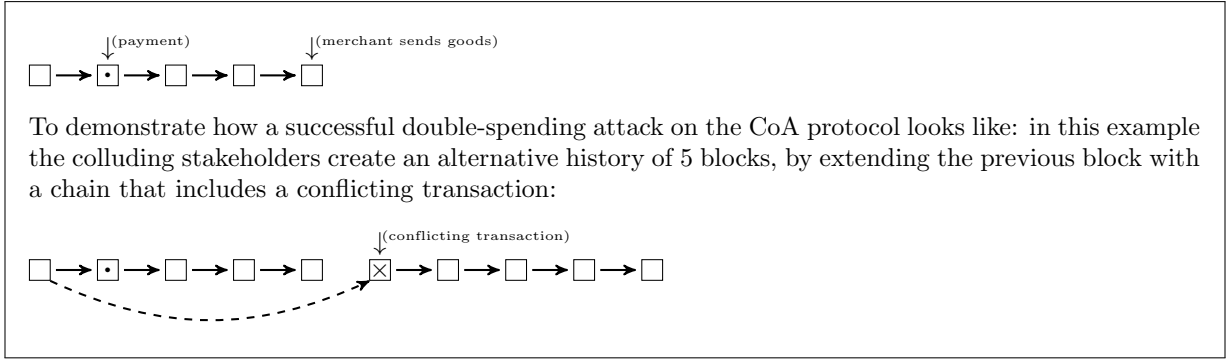

Figure 2: Illustration of a double-spending attack in the CoA system.

### 2.2.2 Rational collusions

Stakeholders may wish to collude and skip the last several blocks as if they did not exist, i.e., to extend the blockchain from an earlier block, in order to gain the fees that went to previous stakeholders. This can be mitigated by including in each transaction the index of the latest block that the user who made this transaction is aware of. For example, if the last block of the chain is $B_i$ and it contains a transaction $tx_0$ that specifies that block $i-1$ exists, and a new transaction $tx_1$ that specifies that block $i$ exists is broadcasted, then the stakeholder who creates $B_{i+1}$ cannot reverse $B_i$ to collect the fees of both $tx_0$ and $tx_1$, because $B_i$ must exist in the chain that contains $tx_1$.

The user can even specify in her transaction the index of the block that is currently being created, but this implies that the user will need to send another transaction in the case that the current stakeholder is offline. To mitigate the risk of a double-spending attack, this should be in addition to specifying the lastest block that was already created (cf. Claim 1 for details).

Stakeholders diminish the long-term value of their stake if they engage in such attempts to extract unwarranted fees, as the cryptocurrency then becomes less useful as a medium of exchange. Hence, this strategy is not necessarily rational.

It is also possible to reward stakeholders via monetary inflation and have the transaction fees destroyed to provide a counterbalance, though bribe attacks would then become more likely (cf. Claim 1).

### 2.2.3 Bribe attacks on CoA

Suppose that the number of blocks that merchants consider to be secure against double-spending attacks is $d$, i.e., a merchant will send the goods after she sees that the payment transaction that she received in block $B_{i_1}$ has been extended by $B_{i_2}, B_{i_3}, \ldots, B_{i_d}$ extra blocks. An attacker can now offer bribes to $d+1$ or more stakeholders, for example to the next $i_d+1, i_d+2, \ldots, i_d+d+1$ stakeholders so that they would extend the blockchain starting from the block that preceded $B_{i_1}$ and exclude that payment transaction. The attacker will need to bribe more than $d+1$ stakeholders if some of them refuse the bribe. Since rational stakeholders will not participate in the attack without an incentive, the cost of the attack is at least $\mu(d+1)$ where $\mu$ is the average bribe amount that is given to each stakeholder.

Observe that $\Pr[\{\text{successful attack}\}] < 1$ since some of the stakeholders might be altruistic, some of the rational stakeholders may think that it would be unprofitable to participate in such attacks, and the attacker's funds are not unlimited. Hence, a rational stakeholder will choose to accept the bribe by weighing whether $(\mu + F') \cdot \Pr[\{\text{successful attack}\}] > F \cdot (1 - \Pr[\{\text{successful attack}\}])$, where $F$ and $F'$ are the fee amounts that this stakeholder will collect on the honest chain and the attacker's chain, respectively. Note that $F' = 0$ is likely when the safety mechanisms of Section 2.2.2 are deployed, since it is rational for users to continue to transact on the honest chain as long as the attacker's chain is inferior. Overall, the attacker may need to spend substantially more than $\mu(d+1)$ coins for the attack to succeed.

In Figure 2 we illustrate the nature of a double-spending bribe attack.

Notice that the attacker cannot simply bribe the stakeholders who generated the blocks $B_{i_1}, B_{i_2}, B_{i_3}, \ldots, B_{i_d}$ to create an alternative history of length $d$ in a risk-free manner, as their coins will be confiscated if they double-sign. Hence, the above stands in stark contrast to Section 2.1.2, as the short-term dominant strategy of the PPCoin stakeholders is to participate in the attack, unlike the CoA stakeholders who will forfeit their reward $F$ in case the attack fails. The underlying reason behind this is that in CoA the identity of the next eligible stakeholder is fixed, while PPCoin allows for certain leniency (cf. Section 2.1) in the timing in which a stakeholder is eligible to create a block, which implies that it is problematic to impose a punishment for double-signing in PPCoin. That is, an honest PPCoin stakeholder who first signed a shorter chain and then became aware (due to propagation delays) and signed a longer chain should not be penalized, as it is a low-probability event to become eligible to create a block and earn the reward, and therefore punishments due to honest mistakes (or due to malicious scheduling by competing stakeholders) are troublesome.

Formally, let us restrict ourselves to a limited strategy space (cf. [17]) in which players have to choose one of only these two actions (**),

1. Follow the protocol honestly by signing a block that extends the longest known chain.

2. Accept bribe and sign the attacker's block which extends the secretive chain that the attacker builds.

This restriction can be justified under plausible assumptions. In particular, the $C_0$ penalty can be assumed to be high enough to make the action of double-signing unappealing. This requires the presupposition that the double-signing punishment mechanism is effective in the sense that the evidence of double-signing will be recorded on every fork, and hence the utility of a player is the value of attacker's bribe minus the loss of her $C_0$ security deposit. This also implies that our analysis here only covers forks that are shorter than the $T_0$ deposit duration, in Section 3 we discuss attacks that involve longer forks.

Our objective is to show that the honest strategy is dominant. In fact, we will show that under further assumptions no attack will be initiated, thus only the honest action will be available to the players.

To analyse what merchants can consider to be an appropriate confidence level for security against double-spending in the CoA system, let us make a reasonable assumption regarding the participation rate of stakeholders in the CoA network.

**Density assumption.** Let $\rho > 1/2$. In the longest blockchain, for every segment of $K$ or more potential blocks, at least $\rho K$ of those blocks were created.

While this is a simplifying assumption, it is indeed reasonable, as our presupposition for the CoA network is that its security is derived from stakeholders' participation. Notice that we do not assume that the majority of stakeholders are altruistic (i.e., follow the CoA protocol even if it is against their self-interest). Although an altruistic majority would facilitate a system with better security, a rational majority is far more likely to capture reality.

Let $B_0$ be a block in which some particular payment transactions resides. Let $\delta$ denote the amount missing blocks in largest segment with participation rate $\leq 1/2$ prior to $B_0$, and let $\rho'$ denote the density of the longest segment that follows $B_0$.

In the illustration below, $\delta = 3$ and $\rho' = 10/14$.

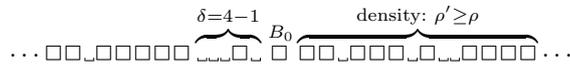

**Claim 1.** Let $\varepsilon$ be the average fee amount that a stakeholder earns for creating a block. Assume that stakeholders are restricted to the strategy space (**). Assume that reversing $B_0$ has a value of $V$ coins to the attacker. If the attacker is rational in the sense that she does not wish to lose coins, then the merchant is safe by waiting until $S$ blocks extend $B_0$ before sending the merchandise, for $S$ that satisfies $V < \varepsilon(\rho'S - \delta + 1)$.

**Proof.** By using the safety extension that is described in Section 2.2.2, we may consider the blocks in a hostile competing fork to be void of transactions, and therefore it is rational of each colluding stakeholder who could otherwise earn $\varepsilon$ coins to demand a bribe of more than this amount. There exist $(1 - \rho')S + \delta$ stakeholders who can contribute to the attack and have already forfeited their turn to create a block, thus the merchant may assume that in the worst case they will collude with the attacker for free. As the other $S + 1$ stakeholders need to be bribed with $\varepsilon$ coins each, $V < \varepsilon(S - (1 - \rho')S - \delta + 1) = \varepsilon(\rho'S - \delta + 1)$ implies that the attack is unprofitable. □

The above argument gives only a crude bound, since it does not capture all the relevant aspects w.r.t. the attack. In particular, the coins that the attacker recovers (in the case of a successful attack) may have less purchasing power, because the cryptocurrency system becomes less valuable whenever double-spending attacks take place. We defer an analysis with more rigorous modeling to future work, as such an analysis will also be useful for Bitcoin and other *Proof of Work* based cryptocurrencies.

**Claim 2.** If the density assumption holds in addition to the assumptions of Claim 1, then the merchant can be confident that it is irrational to carry out a double-spending attack after $B_0$ has been extended by $S$ blocks, for $S$ that satisfies $V < \varepsilon(\rho S - K + 1)$.

**Proof.** According to the density assumption, it holds that $K > \delta$, and since the merchant waited until more than $K$ blocks extend $B_0$ it also holds that $\rho' \geq \rho$. Therefore, $(1 - \rho)S + K \geq (1 - \rho')S + \delta$, and the result follows from Claim 1. □

To get a better sense of things, let us substitute concrete numbers for the above parameters. Suppose for example that $\rho = 7/10$, $K = 20$, $\varepsilon = 10$ coins, and $V = 100$ coins. Hence $10 \cdot (7/10 \cdot S - 19) > 100$ implies that $S = 42$ blocks are sufficient. This means that the merchant will need to wait $\leq 42 \cdot 5$ minutes or 3.5 hours before sending the merchandise, in case CoA is parameterized according to $G_0 = 5$ minutes. As noted in Section 2.2, CoA can indeed accommodate a block interval time that is relatively shorter than that of Bitcoin, hence $G_0 = 5$ minutes is a reasonable supposition.

### 2.2.4 Majority takeover

Consider some stakeholders $A_1, A_2, \ldots, A_m$ who control all of the first $\ell$ locations in the current round. Suppose that these $m$ stakeholders possess $p$-fraction of the *total* stake, and they wish to collude and control all the locations in all of the next rounds, thereby creating a winning chain that consists of *only* their blocks. While this strategy may be irrational as it diminishes the value of their stake, perhaps the $m$ stakeholders prefer a competing system and wish to destroy CoA.

Due to interleaving (cf. Section 2.2), the starting condition for this attack is more difficult to achieve, as these $m$ stakeholders need to control $2\ell$ locations, but let us disregard that for the purpose of in this analysis.

Suppose that $q$-fraction of the honest stake is offline, hence the $m$ stakeholders can give on average a head start of $(\frac{1}{(1-p)(1-q)} - 1)\ell$ blocks to a competing group in each round. Denote $\hat{q} \triangleq (\frac{1}{(1-p)(1-q)} - 1)$. Let $Y$ be the random variable that counts how many of the first $(2 + \hat{q})\ell$ locations of the next round will be controlled by the $m$ stakeholders, so $E[Y] = (2 + \hat{q})\ell p$. Using tail inequality, it holds that

$$\Pr(Y > \ell) = \Pr(Y > \frac{1}{(2 + \hat{q})p} E[Y]) \leq \exp\{-(\frac{1}{(2 + \hat{q})p} - 1)^2 \cdot (2 + \hat{q})\ell p \frac{1}{3}\}.$$

Thus, the amount of hash invocations that these $m$ stakeholders need to compute tends toward infeasibility when $p$ is smaller or when $\ell$ is larger. For example, with $\ell = 459, p = 1/10, q = 1/5$, the $m$ stakeholders will need more than $e^{371} \approx 2^{535}$ hash attempts on average.

Compared with Bitcoin, in a *Proof of Stake* based system such as CoA it is less reasonable to assume that a large combined stake is an hostile external attacker (see [4, Section 2.1]), hence $p$ is likely to be small.

## 2.3 The Dense-CoA pure *Proof of Stake* variant

The Dense-CoA pure *proof of stake* protocol is an alternative variant of CoA in which the identities of stakeholders who should create the next blocks are not known far in advance, with the objective of making collusions and bribe attacks more difficult. Another plus point of Dense-CoA is that it makes it more difficult for rational stakeholders to obtain disproportional rewards. The disadvantages of the Dense-CoA protocol are susceptibility to DoS attacks by large stakeholders, and greater communication and space complexities.

In Dense-CoA, each block is created by a group of $\ell$ stakeholders, rather than by a single stakeholder:

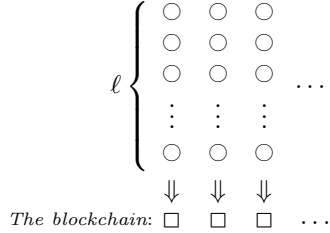

Let $h : \{0,1\}^n \to \{0,1\}^n$ be a one-way permutation. Let us assume for a moment that the block $B_{i-1}$ is associated with a seed $S^{B_{i-1}}$ that was formed by the $\ell$ stakeholders who created $B_{i-1}$. Now, the identity of the stakeholder $A_\ell$ who determines which transactions to include in a block $B_i$ is derived by invoking *follow-the-satoshi* with $hash(i, \ell, S^{B_{i-1}})$ as input, and the identities of the rest of the stakeholders $A_1, A_2, \ldots, A_{\ell-1}$ who must participate in the creation of $B_i$ are derived by invoking *follow-the-satoshi* with $hash(i, j, S^{B_{i-1}})$ for $j \in \{1, 2, \ldots, \ell-1\}$. These $\ell$ stakeholders engage in a two-round protocol to create the current block $B_i$:

- In round 1, for every $j \in \{1, 2, \ldots, \ell\}$, the stakeholder $A_j$ picks a random secret $R_j \in \{0,1\}^n$, and broadcasts $h(R_j)$ to the network.

- In round 2, for every $j \in \{1, 2, \ldots, \ell-1\}$, the stakeholder $A_j$ signs the message $M \triangleq h(R_1) \circ h(R_2) \circ \cdots \circ h(R_\ell)$, and broadcasts her signature $\mathsf{sign}_{sk_j}(M)$ and her preimage $R_j$ to the network.

We require Dense-CoA to use a signature scheme with multisignature [5, 14, 19, 21] support, therefore $A_\ell$ can aggregate the signatures $\{\mathsf{sign}_{sk_j}(M)\}_{j=1}^{\ell}$ into a single signature $\hat{\mathsf{s}}(M)$. Note that the size of $\hat{\mathsf{s}}(M)$ depends only on the security parameter of the signature scheme (and not on $\ell$), and the verification time is faster than verifying $\ell$ ordinary (ECDSA) signatures.

Hence, the stakeholder $A_\ell$ signs and broadcasts a block $B_i$ that consists of the (Merkle root of the) transactions that she wishes to include, the hash of the previous block $B_{i-1}$, the current timestamp, the index $i$, the $\ell$ preimages $R_1, R_2, \ldots, R_\ell$, and $\hat{\mathsf{s}}(M)$. To verify that the block $B_i$ is valid, the network nodes invoke $h$ to compute the images $h(R_1), h(R_2), \ldots, h(R_\ell)$, then concatenate these images to form $M$, and then check that $\hat{\mathsf{s}}(M)$ is a valid signature of $M$ with respect to the public keys $pk_1, pk_2, \ldots, pk_\ell$ that control the winning satoshis of the stakeholders $A_1, A_2, \ldots, A_\ell$.

The seed $S^{B_i}$ is defined as $hash(R_1 \circ R_2 \circ \cdots \circ R_\ell)$. Notice that $S^{B_i}$ is computationally indistinguishable from random even if only a single stakeholder $A_j$ picked a random $R_j$, under the assumption that $n$ is sufficiently large so that the OWP $h$ is resistant to preimage attacks.

If some of the $\ell$ stakeholders are offline or otherwise withhold their signatures, then after $G_0$ time the nodes who follow the protocol will set $t = 1$ and derive alternative $\ell$ identities from the previous block $B_{i-1}$, by invoking *follow-the-satoshi* with inputs $hash(i, t\ell + j, S^{B_{i-1}})$ for $j \in \{1, 2, \ldots, \ell\}$. The starting index $t\ell + j$ should be specified in the new block $B_i$ so that the verification of blocks will be simpler, and the gap between the timestamps of $B_{i-1}$ and $B_i$ must be at least $tG_0$. As with CoA, the honest nodes consider the blockchain with the largest amount of valid blocks to be the winning blockchain, and disregard blocks with a timestamp that is too far into the future relative to their local clock.

The parameters $C_0, C_1, T_0$ of the CoA protocol (cf. Section 2.2) are utilized by the Dense-CoA protocol in exactly the same way.

The parameter $\ell$ should be big enough in order to resist large stakeholders from controlling consecutive seeds $\{S^{B_i}, S^{B_{i+1}}, \ldots\}$ and re-deriving themselves. For example, to force a stakeholder who holds 5% or 10% of the total stake into making $\approx 2^{100}$ *hash* invocations on average until re-deriving herself as all of the $\ell$ identities of the next block, we need $\ell = 23$ or $\ell = 30$, respectively. However, if we set $G_0 = 5$ minutes and $\ell = 23$, a malicious stakeholder with e.g. 10% of the total stake will have $1 - (90/100)^{23} \approx 91\%$ probability to be one of the derived stakeholders $A_1, A_2, \ldots, A_\ell$ and then refuse to participate in creating the next block, hence it will take $5 \cdot (1 - 91\%)^{-1} \approx 56$ minutes on average to create each next valid block while this attack is taking place (actually less than 56 minutes because chains that extend blocks prior to the last block can also become the longest valid chain).

Overall, the main difference between the Dense-CoA and CoA protocols is that Dense-CoA offers improved security over CoA in terms of double-spending attacks, but weaker security against DoS attacks by large stakeholders who wish to harm the cryptocurrency. Also, Dense-CoA prevents a rational stakeholder from influencing the seed in an attempt to earn more rewards than her fair share, unless she colludes with all the other $\ell - 1$ stakeholders who create the next block. The Dense-CoA protocol is less efficient than CoA due to the preimages $R_1, R_2, \ldots, R_\ell$ that need to be stored in each valid block, and the two-round protocol that requires a greater amount of network communication to create each successive block.

# 3 Solidification of the ledger history

Any decentralized cryptocurrency system in which extending the ledger history requires no effort entails the danger of costless simulation [25], meaning that an alternative history that starts from an earlier point of the ledger can be prepared without depleting physical resources and hence without a cost. This is a problem because a rational adversary who has little or no stake in the system may try to attack by replacing an arbitrarily long suffix of the current ledger history with an alternative continuation that benefits her. Further, a malicious adversary who does not operate out of self-interest is also more likely to attempt this kind of an attack, as she would not incur a monetary loss for executing the attack.

In the case of pure *Proof of Stake* systems, this danger can manifest itself in the following form. Consider participants who held coins in the system a long time ago and have since traded those coins in exchange for other goods, so they are no longer stakeholders of this system. These participants can now collude to extend the ledger from the point at which they had control over the system, and it may indeed be rational for them to mount this attack because it is costless and would have no detrimental outcome from their standpoint, as they have no stake in the current system.

More specifically, let us examine how this attack looks like in the CoA or Dense-CoA systems. Even a single stakeholder with few coins can fork the blockchain and create an alternative branch with large enough time gaps as she re-derives herself to create subsequent blocks, but according to the timestamp rules for valid blocks, the other participants will reject this alternative branch (even though it contains more blocks) because the timestamps will be too far ahead in the future relative to their local time. Therefore, if the average participation level among current stakeholders is $p\%$, and the stakeholders who collude to carry out this attack have had control at the earlier history over $q\%$ of the coins, then $q > p$ implies that the attack will succeed. Because $p\% = 1$ is highly unlikely, and collusion among participants who held $q\% > p\%$ stake at an earlier point is costless and rational, this attack vector appears to be quite dangerous.

To mitigate this attack, we propose periodic checkpointing as a rigid protocol rule that extends the CoA and Dense-CoA protocols, as follows:

- Denote by $T_0 = 2T_1$ the double-spending safety bound of Section 2.2.

- The blocks at gaps of $T_1$ are designated as *checkpoint* blocks: the genesis block is a checkpoint block, and any block that extends a checkpoint block by exactly $T_1$ additional blocks is a candidate checkpoint block.

- When a node that follows the protocol receives for the first time a candidate checkpoint block $B_j$ that extends the candidate checkpoint block $B_i$ such that $j = i + T_1$ (or $j > i + T_1$ if some stakeholders were inactive), she solidifies $B_i$ meaning that she disallows any changes to the history from the genesis block until $B_i$, though $B_j$ can still be discarded as a result of a competing fork.

Since the double-spending safety bound is $T_0$, a stakeholder who creates a block can spend the coins only after an intermediate checkpoint block is already solidified, so the costless simulation threat is mitigated (if $C_0$ is substantial). See Figure 3 for an illustration.

However, this checkpointing mechanism presents two significant problems:

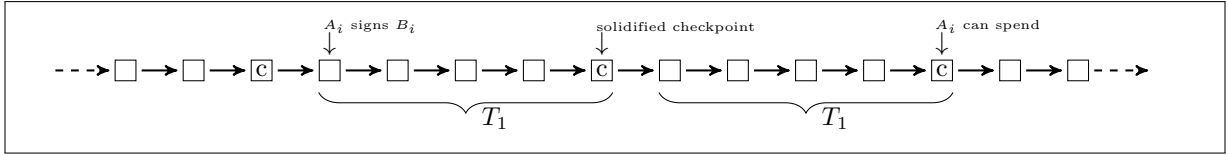

Figure 3: Checkpoint solidification prior to the $T_0 = T_1 + T_1$ bound.

1. New nodes who enter the decentralized network for the first time cannot tell whether the checkpoint blocks that they receive are trustworthy.

2. Due to propagation lag, adversarial stakeholders can collude by preparing an alternative branch of length $T_1 + 1$, and broadcast the competing forks at the same time, thus creating an irreversible split among the network nodes.

The first problem needs to be handled by utilizing a "Web of Trust" type of mechanism that is external to the cryptocurrency system. This means that participants who are unaware of the current state of the system should rely on reputable sources to fetch the blockchain data up to the latest checkpoint.

The second problem should also be resolved manually, meaning that participants who become aware of a network split can decide to instruct their node to switch to the other faction, e.g. if they see that they are in the minority. Note, however, that the second problem becomes increasingly unlikely for larger $T_1$ values. The exemplary parameters that we proposed in section 2.2 imply that a fork of $T_1 + 1$ blocks represents more than one week of ledger history.

Let us mention that PoW based systems such as Bitcoin also benefit from having checkpoints, as it helps to bootstrap new nodes who enter the network in a more efficient manner. The Bitcoin checkpoints also thwart DoS attacks that would extend the genesis block with low-difficulty large blocks, although "headers-first" client behavior[3] can be sufficient for mitigating DoS, as the client will quickly discard an alternative branch that is being transmitted unless its cumulative PoW outcompetes the current best branch. Both CoA and Bitcoin can utilize SNARK (see e.g. [13]) to represent the ledger state (a.k.a. the set of unspent coins) that is reached by a checkpoint, as sending a SNARK to new nodes is more efficient than sending the entire ledger history. While CoA checkpoints rely on an element of trust, Bitcoin can take advantage of SNARK checkpoints that are trust-free in the sense that a new node will have exactly the same level of confidence in the ledger state as the existing nodes, because the SNARK will prove that the protocol rules have been followed from the genesis block until the current state.

## 4 Issuance of the money supply

With a cryptocurrency that does not incorporate a PoW component, one straightforward way to distribute the money to the interested parties is by having an IPO or an auction of some sort. However, this implies that initially the money supply is controlled by a central party, which means that it is less likely that the cryptocurrency can be decentralized. The concerns that arise with this approach are relevant not only to *Proof of Stake* based cryptocurrencies, since it is always the case that large stakeholders can manipulate the market more easily.

We propose to use PoW only for the initial issuance of the coins that should then circulate in the cryptocurrency system, in a way that pegs the value of the newly minted coins to the cost of producing these coins (cf. [31]). This can be done by dictating that the cost of producing a coin in terms of electricity and erosion of the equipment will be approximately fixed throughout the issuance process, and hence:

- If the value of each coin is more than the cost of producing a coin, then more mining equipment will be

---

[3] See https://bitcoin.org/en/glossary/headers-first-sync.

brought online to produce larger amounts of coins at the fixed cost, and then larger amounts of newly minted coins will come into existence - which implies that the value of each coin decreases.

- If the value of each coin is less than the cost of producing a coin, then some of the mining equipment that participates in the minting process will quit, and then smaller amounts of coins will come into existence - which implies that the value of each coin increases.

In order to have a fixed cost of production, we simply need to remove the difficulty readjustment mechanism that Bitcoin based systems use. This means that there will no longer be a predicable gap of $x$ minutes on average (e.g. $x = 10$ with Bitcoin) between the PoW blocks that are being created, and instead the gap between PoW blocks will directly correspond to the amount of mining power that participates in the block creation process. This way, if an individual miner could create a block once every $y$ minutes on average in accord with the fixed PoW difficulty, then she will still be able to create blocks once every $y$ minutes even if additional mining power joins the process, though her blocks will represent a smaller portion of the total amount of blocks that are being created. To avoid the possibility that blocks get generated extremely fast in the case that the cryptocurrency continues to have a high market value even when many newly minted coins come into existence, the protocol can still specify a difficulty readjustment rule that imposes a minimal average gap of e.g. 1 minute between PoW blocks. The minimal gap is desirable as otherwise there can be many orphaned blocks, which amplifies the phenomenon where miners who are well-connected to high concentrations of the mining power receive more rewards than what their proportion relative to the total mining power should imply.

Bitcoin's difficulty readjustment mechanism is different than the above mechanism because the value of a coin is ultimately determined by long-term fundamentals that are derived from the adoption level of the cryptocurrency among merchants, rather than the amount of miners who wish to mint new coins during a certain time period. Hence, given the current market value of a coin, the cost of minting new coins in the Bitcoin system is determined according to how many miners currently participate in the production process, i.e., a greater participation level implies that an individual miner will be rewarded with a smaller proportion of the fixed amount of coins that is being produced every 10 minutes of average. When the difficulty readjustment mechanism is not deployed, the cost of production for each individual miner is not affected by the overall mining power that participates in the production process.

Thus, the advantage of the coins distribution mechanism that CoA supports is that it allows for a stable exchange rate during the inflationary phase of the cryptocurrency, unlike the quite dramatic price fluctuations of Bitcoin.

For the CoA system to deploy such a PoW issuance scheme, the overall protocol should specify that the CoA network nodes are allowed to create transactions that spend newly minted coins from the PoW blockchain if and only if the minted coins are buried behind a sufficiently large amount $n$ of PoW blocks. In Bitcoin $n = 100$, but here $n > 100$ may be prudent, in case the fixed PoW difficulty becomes relatively easier for future mining hardware. If the minimal average gap rule is in place, then a sensible value for $n$ can be specified with no uncertainty issues. For a cryptocurrency without infinite monetary inflation, the CoA network nodes can stop listening for new PoW blocks after the last PoW block is created (for example the last spendable PoW block can be solidified by making the PoW blocks that follow it unspendable), as well as completely discard the PoW blockchain after all the minted coins have been spent.

Let us note that an alternative approach for the initial issuance of the coins is "one-way pegging" a.k.a. "proof of burn" (see [6,7]), by which coins of an earlier cryptocurrency are destroyed in exchange for newly minted coins in the system that is being launched. Under the assumption that the earlier cryptocurrency is decentralized, this issuance method is indeed decentralized as well. However, this issuance method does not imply that the value of minted coins in the new system is pegged to the cost that it took to produce these coins.

# 5 Conclusion

It is challenging to design sustainable decentralized cryptocurrency protocols that do not rely on depletion of physical scarce resources for their security maintenance. Our analysis argues that the security of existing such protocols is lacking. We offer novel constructions of pure *Proof of Stake* protocols that avoid depletion of physical scarce resources, and argue that our protocols offer better security than existing protocols. Future work could extend the scope of our analysis to broader strategy spaces.

**Acknowledgments.** We thank Gregory Maxwell and Andrew Miller for insights regarding the costless simulation attack.